\documentclass[lettersize,journal]{IEEEtran}
\usepackage{amsmath,amsfonts}
\usepackage{algorithmic}
\usepackage{array}
\usepackage[caption=false,font=normalsize,labelfont=sf,textfont=sf]{subfig}
\usepackage{textcomp}
\usepackage{stfloats}
\usepackage{url}
\usepackage{verbatim}
\usepackage{graphicx}
\usepackage{hyperref}
\usepackage{nicefrac, xfrac}
\usepackage{xurl}
\usepackage{csquotes}
\usepackage{balance}

\def\BibTeX{{\rm B\kern-.05em{\sc i\kern-.025em b}\kern-.08em
    T\kern-.1667em\lower.7ex\hbox{E}\kern-.125emX}}

\begin{document}
\title{Throughput and Delay Performance of Slotted Aloha in SmartBANs under Saturation Conditions}
\author{Anastasios C. Politis, \IEEEmembership{Senior Member, IEEE} and Constantinos S. Hilas, \IEEEmembership{Senior Member, IEEE}
\thanks{A. C. Politis, corresponding author, and C. S. Hilas are with the Department of Computer, Informatics and Telecommunications Engineering, International Hellenic University (Serres Campus), Serres, Greece (e-mail: anpol@ihu.gr; chilas@ihu.gr) \\ $\copyright$ 2024 IEEE. Personal use of this material is permitted. The final authenticated version is available at: https://doi.org/10.1109/LNET.2024.3467031.}}

\markboth{Journal of \LaTeX\ Class Files,~Vol.~XX, No.~XX, September~XXXX}%
{How to Use the IEEEtran \LaTeX \ Templates}

\maketitle

\begin{abstract}
This letter evaluates the performance of the slotted Aloha protocol defined by the European Telecommunication Standard Institute (ETSI) SmartBAN specification, under saturation conditions. For this purpose, we develop a two-dimensional Discrete Time Markov Chain (DTMC) to model the operational details of the protocol and assess its performance in terms of saturation throughput and average end-to-end delay. The accuracy of the proposed model is validated by means of simulation which reveals a very good match among theoretical and simulation results. The model can be used for protocol performance prediction and optimization purposes.
\end{abstract}


\section{Introduction}
\IEEEPARstart{W}{ireless} Body Area Networks (WBANs) are designed to operate on or within the human body. Their typical use-cases include health monitoring for well being or sporting activities and extend to military and rescue operations \cite{b1,b2}. The typical architecture of a WBAN involves a number of in or on-body sensors that communicate with a central node (coordinator or hub) to deliver their recorded data.

The IEEE 802.15.6 \cite{b3} and the ETSI SmartBAN \cite{b4} are two of the most popular standards that define the functional characteristics of WBANs at the Physical (PHY) and Medium Access Control (MAC) layers. Both standards define a Slotted Aloha Channel Access (SACA) mechanism that allows a random access to the wireless medium under specific transmission rules. The MAC-layer of SmartBANs is designed to support configurable Quality of Service (QoS) provisioning by offering four \emph{User Priorities (UPs)} \cite{b5}. A node can have its traffic assigned to one of the available $UP$s which are related to specific MAC-layer parameters that influence the node's probability to access the medium \cite{b6}. Thus, traffic differentiation can be achieved in different use-cases. A collection of typical use-cases of ETSI SmartBANs can be found in \cite{b7}.

According to the SmartBAN specification, the primary scope of SACA is to be used for management frames transmission during the Control and Management (C/M) period. Nevertheless, SACA can also be exploited by the participating nodes to transmit data frames. Hence, a saturation throughput analysis of the SACA period will provide valuable insights on its maximum capability to deliver network traffic. While such studies are available in the literature for the IEEE 802.15.6 standard \cite{b8, b9}, an equivalent investigation has not been detected for the ETSI SmartBANs. Some MAC-layer performance related studies for SmartBAN technology exist \cite{b10, b11}, but none of them address SACA. Furthermore, besides being an unexplored issue, examining the protocol's peak performance will provide a reference point from which optimization efforts can be initiated.

In this paper, the saturation throughput performance of the slotted Aloha version, as specified by the ETSI SmartBAN standard, is investigated and is accompanied by a delay analysis. Based on the renowned work of G. Bianchi in performance modeling of MAC protocols in wireless networks \cite{b12}, we illustrate the operational specifics of the SACA period with the aid of a two-dimensional DTMC, which enables a classical Markovian analysis. This leads to an expression for the stationary transmission probability that a SmartBAN node transmits in a randomly chosen time slot, which is, thence, used to provide an estimation of the saturation throughput per $UP$. Furthermore, the proposed DTMC effectively captures the delay dynamics of the SACA period allowing the evaluation of the average delay experienced by packets belonging to different traffic categories. The model's capability to deliver credible results is validated via a customized simulator specifically developed for this purpose. According to our literature research, no study exploiting a Markovian analysis to model the operation of SACA in ETSI SmartBANs has been discovered. The proposed model can be used to assess the system's  capability to deliver satisfactory QoS levels under different use-cases. Moreover, it can be used for fine-tuning the MAC-layer parameters of the protocol in order to optimize its performance. 

The rest of this letter is structured as follows. In Section II an overview of the SACA, as defined by the ETSI SmartBAN standard, is provided. In Section III, the considered system model is described and Section IV presents the proposed analysis. Section V includes the numerical and simulation results. The paper is concluded with Section VI with concluding remarks.

\section{Overview of the SACA in ETSI SmartBAN}

The slotted Aloha version in SmartBAN differs from the classical slotted Aloha protocol since it incorporates a transmission probability reduction mechanism based on the number of failed transmission attempts. 

A node that has a data or a management frame ready for transmission, can initiate a SACA based on the value of \emph{Contention Probability (CP)}, which represents the likelihood of that node accessing the medium (i.e., transmitting) at the start of a given time slot \cite{b6}. $CP$ is denoted in this work as $\alpha$. Probability $\alpha$ can take discrete values in the range $[CP_{min},CP_{max}]$. The values of $CP_{min}$ and $CP_{max}$ depend on the $UP$ of the node's traffic, as defined by the standard. 

A node with a frame ready for transmission, waits for the start of a time slot and chooses its $\alpha$ value according to the following rules \cite{b5}:

\begin{itemize}
\item{if this is the node's first SACA attempt or its last SACA attempt was successful, then $\alpha=CP_{max}$}.
\item{if the node's last SACA attempt was unsuccessful:}
\begin {itemize}
\item{if the total number of unsuccessful attempts hitherto is an even number and $\alpha\geq 2\times CP_{min}$ holds, then $\alpha=\sfrac{\alpha}{2}$. Otherwise, $\alpha$ retains its current value.}
\item{if the total number of unsuccessful attempts hitherto is an odd number, then $\alpha$ remains unchanged.}
\end{itemize}
\end{itemize} 

The standard dictates the emission of a positive (ACK) or a negative (NACK) acknowledgment frame from the receiving device based on the acknowledgment policy indicated in the received data frame. Selecting the NACK policy will trigger the transmission of a NACK frame only when the data frame is received with errors. Otherwise, no acknowledging frame will be produced by the receiver.

\section{System Model}
In this work, a SmartBAN system with $n$ number of nodes is assumed. Note that the standard indicates that the maximum number of nodes supported in a SmartBAN is set to 16 but recommends a system with 8 stations. The nodes are connected to the coordinator (hub) in a star topology. The NACK acknowledgment policy is considered and the channel is assumed error-free (an assumption that is typically adopted in similar studies \cite{b8}, \cite{b9}, \cite{b12}). Thus, no acknowledgment frames are produced by the receiving device. All nodes always have a data frame ready for transmission  (i.e., they operate in saturation conditions) and their data frames may occupy a single or multiple equally sized time slots. Subsequent fragments of a frame that exceeds the slot length also content for medium access at the start of future time slots. Lastly, we consider the case where every node in the system produces traffic that belongs to the same $UP$ (i.e., a homogeneous network). 

\begin{figure*}[t!]
\centering
\includegraphics[width=1\linewidth]{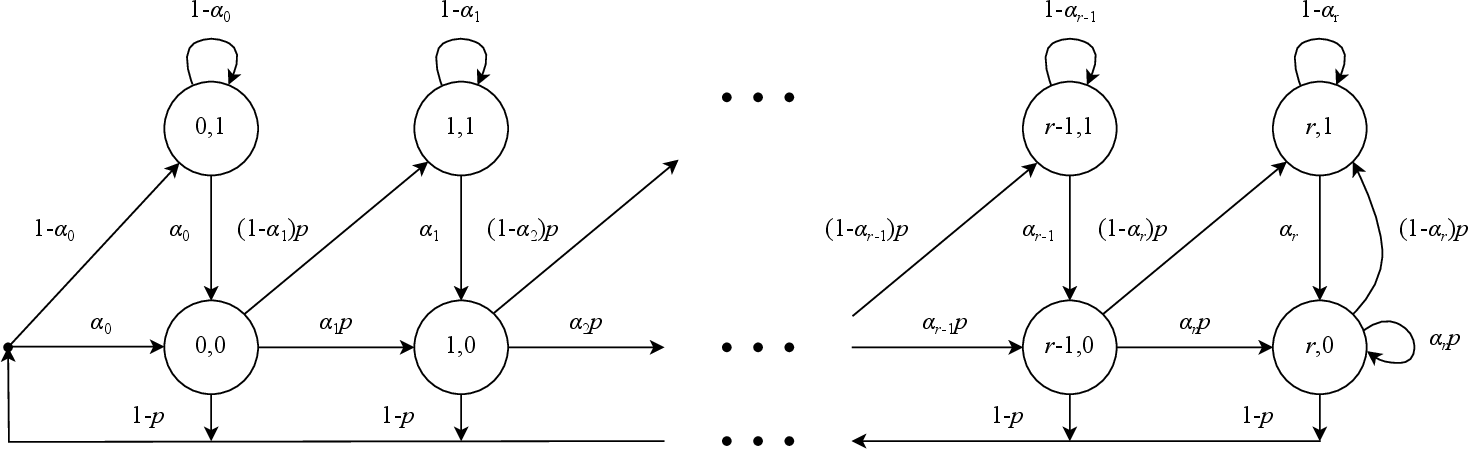}
\caption{The proposed two-dimensional DTMC.}
\label{fig1}
\end{figure*}

\section{Analysis}
Based on the set of operational rules of the ETSI SmartBAN slotted Aloha outlined in Section II, it can be seen that $\alpha$ is set at $CP_{max}$ for the initial transmission of new packet and will reach the $CP_{min}$ value after a specific number of retransmission attempts. For every other retransmission attempt $\alpha$ is halved until it reaches $CP_{min}$. Any new failed retransmission from that point onwards will keep the value of $\alpha$ unchanged. We refer to that point as the $CP_{min}$ stage, $r$. Hence, $r=\sfrac{CP_{max}}{CP_{min}}$. The values of $CP_{min}$, $CP_{max}$ (according to the SmartBAN specification) and $r$ for different $UP$s are summarized in Table \ref{tab1}. 

\begin{table}[t!]
\caption{$CP_{min}$, $CP_{max}$ and $r$ values for different UPs}
\begin{center}
\begin{tabular}{|c|c|c|c|}
\hline
\textbf{User}&\multicolumn{2}{|c|}{\textbf{Contention Probability}}&\textbf{Maximum}  \\
\cline{2-3} 
\textbf{Priority} & $CP_{max}$& $CP_{min}$ &\textbf{Retransmission Stage, $r$} \\
\hline
0 (low)& $\sfrac{1}{8}$&$\sfrac{1}{16}$& 2   \\
\hline
1 (mid)& $\sfrac{1}{4}$&$\sfrac{1}{16}$& 4   \\
\hline
2 (high)& $\sfrac{1}{2}$&$\sfrac{1}{8}$& 4   \\
\hline
3 (emergency)& 1& $\sfrac{1}{2}$& 2  \\
\hline
\end{tabular}
\label{tab1}
\end{center}
\end{table}

Focusing on a saturated node during a randomly selected system slot, that node may be at a particular retransmission stage. Furthermore, that node may be either at a transmitting or waiting (deferring transmission) state. If it is transmitting, the transmission may be successful or unsuccessful. This situation can be modeled as a two-dimensional DTMC. The proposed DTMC to model the slotted Aloha functionality is depicted in Fig. \ref{fig1}. The DTMC consists of a series of states $(i,j)$ with $i\in \{0,1,...,r\}$ and $j\in\{0,1\}$. States $(i,0)$ represent transmission states and states $(i,1)$ represent waiting (deferring) states.

Let us denote with $\tau$ a nodes' probability to transmit a packet in a randomly chosen time slot. Let us, also, denote with $p$ the collision probability of a randomly chosen packet conditioned to the event that is being transmitted, which we consider to be constant and independent of any previous collision instances \cite{b12}. Hence:
\begin{align}
p=1-(1-\tau)^{n-1}.
\label{eq1}
\end{align}

With respect to the DTMC in Fig \ref{fig1}, the one-step transition probabilities are:
\begin{align}
\begin{cases}
P\{k,0|k-1,0\}=\alpha_kp&\text{for $k\in[1,r]$}\\ 
P\{k,0|k,1\}=\alpha_k & \text{for $k\in[0,r]$}\\
P\{0,0|k,0\}=(1-p)\alpha_0 & \text{for $k\in[1,r]$}\\
P\{r,0|r,0\}=\alpha_rp\\
P\{k,1|k-1,0\}=(1-\alpha_k)p & \text{for $k\in[1,r-1]$}\\
P\{k,1|k,1\}=1-\alpha_k & \text{for $k\in[0,r]$}\\
P\{0,1|k,0\}=(1-p)(1-\alpha_0) & \text{for $k\in[1,r]$}\\
P\{r,1|r,0\}=(1-\alpha_r)p.
\end{cases}
\label{eq2}
\end{align}

\noindent The first four equations describe, probabilistically, the ways a node is found in the transmission state of a specific retransmission stage. The next four expressions model the other alternative, i.e., all the cases where a node ends up in the deferral state. 

Given the slotted Aloha rules specified in the standard, probability $\alpha$ in every retransmission stage is given by:
\begin{align}
\alpha_k=\frac{CP_{max}}{2^{\lfloor\sfrac{k}{2}\rfloor}} \qquad \text{for $k\in[0,r]$}.
\label{eq3}
\end{align}

Denoting as $s_{i,j}$ the probability of state $(i,j)$, it is easy to show that for the DTMC shown in Fig. \ref{fig1} the following equations hold for transmission states:
\begin{align}
s_{k,0}=p^ks_{0,0} \qquad \text{for $k\in[1,r-1]$},
\label{eq4}
\end{align}
\begin{align}
s_{r,0}&=\frac{p^r}{1-p} s_{0,0}.
\label{eq5}
\end{align}

For deferring states:
\begin{align}
s_{k,1}=\frac{1-\alpha_k}{\alpha_k}p^ks_{0,0} \qquad \text{for $k\in[1,r-1]$},
\label{eq6}
\end{align}
\begin{align}
s_{0,1}&=(1-\alpha_0)s_{0,1}+(1-\alpha_0)(1-p)\sum_{k=0}^rs_{k,0}\nonumber\\
&=\frac{1-\alpha_0}{\alpha_0}(1-p)s_{0,0}\Bigl[1+p+p^2+...+p^{r-1}+\frac{p^r}{1-p}\Bigr]\nonumber\\
&=\frac{1-\alpha_0}{\alpha_0}s_{0,0},
\label{eq7}
\end{align}
\noindent and
\begin{align}
s_{r,1}&=(1-\alpha_r)s_{r,1}+(1-\alpha_r)ps_{r-1,0}+(1-\alpha_r)ps_{r,0}\nonumber \\
&=\frac{1-\alpha_r}{\alpha_r}\frac{p^r}{1-p}s_{0,0}.
\label{eq8}
\end{align}

Given the normalization condition:
\begin{align}
1&=\sum_{k=0}^rs_{k,0}+\sum_{k=0}^rs_{k,1}\nonumber \\
&=s_{0,0}\Bigl[\frac{1}{1-p}+\frac{1-\alpha_r}{\alpha_r}\frac{p^r}{1-p}+\sum_{k=0}^{r-1}\frac{1-\alpha_k}{\alpha_k}p^k\Bigr]\Rightarrow \nonumber \\
&s_{0,0}=\frac{1}{\frac{1}{1-p}+\frac{1-\alpha_r}{\alpha_r}\frac{p^r}{1-p}+\sum_{k=0}^{r-1}\frac{1-\alpha_k}{\alpha_k}p^k}.
\label{eq9}
\end{align}

Now, the transmission probability, $\tau$, is expressed as:
\begin{align}
\tau&=\sum_{k=0}^rs_{k,0}=\frac{s_{0,0}}{1-p}\nonumber \\
&=\frac{1}{1+\sum_{k=0}^r\frac{1-\alpha_k}{\alpha_k}p^k-\sum_{k=0}^{r-1}\frac{1-\alpha_k}{\alpha_k}p^{k+1}}.
\label{eq10}
\end{align}

We can particularize the above expression for the different $UP$s defined under the SmartBAN standard by expanding the sums in the denominator of (\ref{eq10}), considering the values of $r$ in Table \ref{tab1} and utilizing (\ref{eq3}):
\begin{align}
\tau=
\begin{cases}
\dfrac{CP_{max}}{1+p^2}&\text{for $UP=0$ and $UP=3$}\\ 
\\
\dfrac{CP_{max}}{1+p^2+2p^4}&\text{for $UP=1$ and $UP=2$}.
\end{cases}
\label{eq11}
\end{align}

Equations (\ref{eq1}) and (\ref{eq11}) can be solved to obtain the values of $\tau$ and $p$ for the different $UP$s, which can be used to obtain the probability that a given time slot is occupied with a successful transmission. Now, the normalized throughput, $S$, can be expressed by this probability in order to reflect the fraction of time slots that are effectively utilized for successful transmissions. Hence, as also given by \cite{b8}, :
\begin{align}
S = n\tau (1-p)=n\tau (1-\tau)^{n-1}.
\label{eq12}
\end{align} 

\subsection{Delay analysis} 

Packet delay can be defined as the time required for a packet to be received successfully, starting from the time instant the packet is placed in the head of the transmission queue. This period will include idle and collision slots that the packet will experience until it successfully reaches its destination. The procedure will continue until a maximum number of retransmission attempts is reached, after which, in the case of a collision, the packet gets dropped and no delay is measurable. However, the SmartBAN standard does specify such value and we consider the theoretical case of infinite retransmission attempts.

Based on the DTMC and concentrating on a specific node, the average number of time slots that node spends in deferring a packets' transmission in retransmission stage $i$, until it finally attempts its transmission, is simply $\alpha_i\sum_{j=1}^{\infty}j(1-\alpha_i)^{j-1}=\frac{1}{\alpha_i}$. That transmission may be successfully delivered or it may collide. The former case will signalize the end-to-end delay for that packet. The latter case will lead to the next retransmission stage which will add, on average, $\frac{1}{\alpha_{i+1}}$ time slots to the already accumulated time slots from the previous stage(s). The same cases as before are applicable in this new retransmission attempt. 

Hence, the average delay, expressed in time slots, for a packet belonging to a specific $UP$, is:
\begin{align}
E[D] &= \sum_{i=0}^{r-1}p^i(1-p)\Bigl(\sum_{j=0}^{i}\frac{1}{a_j}\Bigr)\nonumber\\
&+\sum_{i=0}^{\infty}p^{r+i}(1-p)\Bigl[\Bigl(\sum_{j=0}^{r-1}\frac{1}{a_j}\Bigr)+(i+1)\frac{1}{a_r}\Bigr].
\label{eq13}
\end{align}

After some algebraic manipulations and simplifications on (\ref{eq13}), one can particularize the average delay experienced by packets belonging to a specific $UP$ as below:
\begin{align}
E[D]=
\begin{cases}
\dfrac{1+p^2}{CP_{max}(1-p)}&\text{for $UP=0$ and $UP=3$}\\ 
\\
\dfrac{1+p^2+2p^4}{CP_{max}(1-p)}&\text{for $UP=1$ and $UP=2$}.
\end{cases}
\label{eq14}
\end{align}

Alternatively, one can reach (\ref{eq14}) following a more intuitive way. The probability that a specific node ends up with a successful transmission, $P_s$, is simply $\tau(1-p)$. Given the constant and equal size of transmission, collision and idle slots, the average delay experienced by a packet transmitted by that node, expressed in time slots, is mathematically described as:
\begin{align}
E[D] = P_s\sum_{i=1}^{\infty}i(1-P_s)^{i-1}=\frac{1}{P_s}=\frac{1}{\tau(1-p)},
\label{eq15}
\end{align} 

\noindent which leads to (\ref{eq14}) for the available $UP$s.

\subsection{Remarks}

In the case of $\tau=0$, no node transmits any data frame, and if $\tau=1$ when $n>1$ (e.g., by tampering with the SACA rules), transmitted data frames collide with 100\% certainty ($p=1$). In both cases, no throughput is achieved, and no delay is measurable. While (\ref{eq12}) reflects these extreme cases, in terms of saturation throughput, (\ref{eq14}) and (\ref{eq15}) cannot describe them, with regard to packet end-to-end delay. Thus, (\ref{eq14}) and (\ref{eq15}) hold when $0<\tau<1$ for $n>1$ and when $\tau>0$ for $n=1$.


\section{Model validation and results}

Four different scenarios (one per $UP$) were considered with $UP$-homogeneity in the system. In each scenario, the number of participating nodes in the system is increased gradually from one to 16, which is the maximum supported number in a SmartBAN network. For each node population size, a theoretical and a simulation-based value of the normalized saturation throughput and average end-to-end delay are obtained. The corresponding simulation derived values, were produced by a simple customized slotted Aloha simulator which was developed in Python programming language. The simulator was based on a simple classic slotted Aloha simulation program freely available in Github \cite{b13} and whose ability to produce credible results was confirmed. That simulator was retrofitted with the operational details of the slotted Aloha version defined in the ETSI SmartBAN standard. For each simulation run, a total of $10^5$ equally sized slots were considered. This value allowed the simulation program to generate smoother average values. Higher numbers extended the simulation time with no significant benefits.

\begin{figure}[t!]
\centerline{\includegraphics[width=0.8\linewidth]{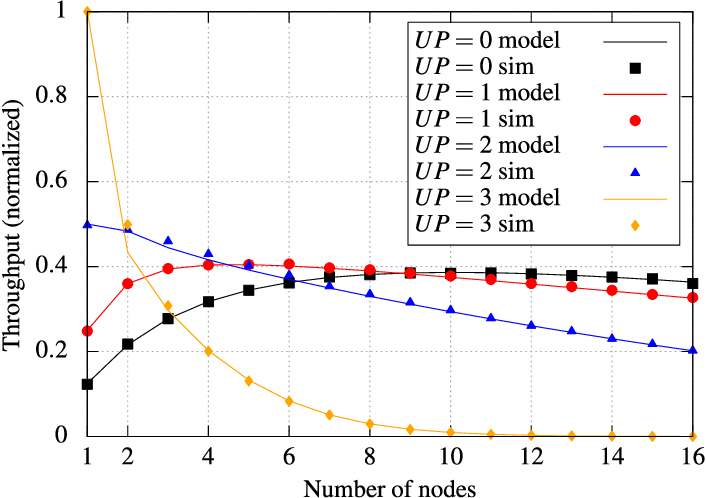}}
\caption{Normalized throughput versus the number of nodes for the different $UP$s.}
\label{fig2}
\end{figure}

\begin{figure}[t]
\centerline{\includegraphics[width=0.8\linewidth]{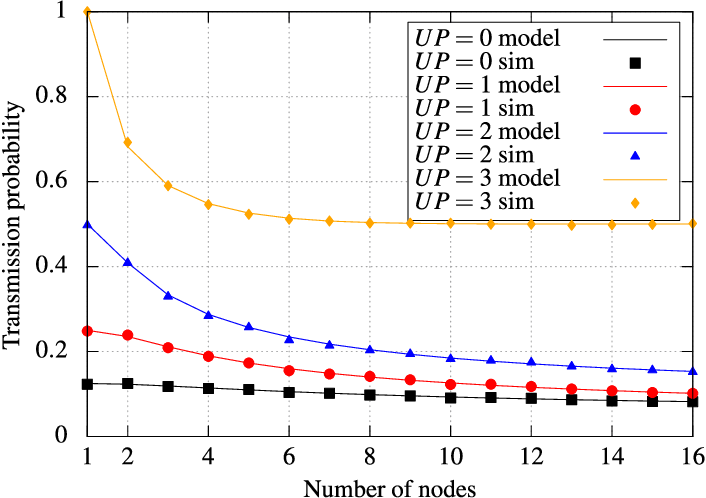}}
\caption{Transmission probability versus the number of nodes for the different $UP$s.}
\label{fig3}
\end{figure}

\begin{figure}[t]
\centerline{\includegraphics[width=0.8\linewidth]{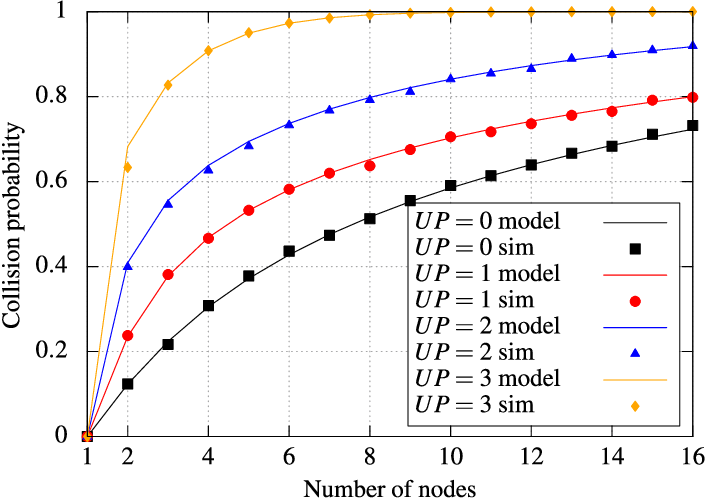}}
\caption{Collision probability versus the number of nodes for the different $UP$s.}
\label{fig4}
\end{figure}

\begin{figure}[t]
\centerline{\includegraphics[width=0.8\linewidth]{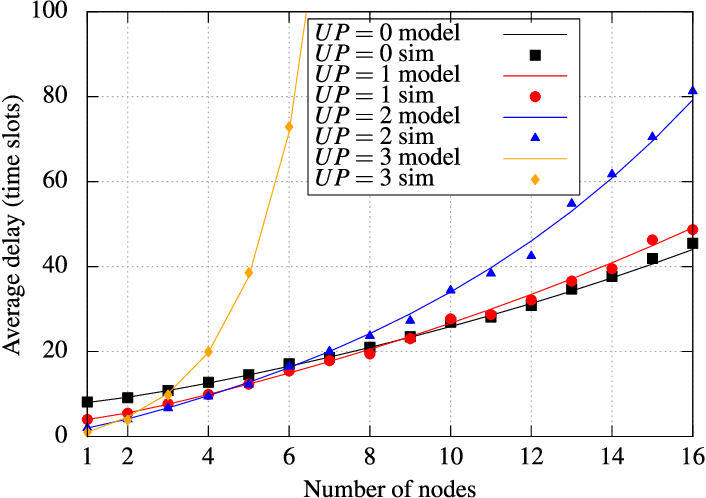}}
\caption{Average end-to-end delay (measured in time slots) versus the number of nodes for different $UP$s.}
\label{fig5}
\end{figure}

The comparison among the theoretic and the simulation-based values of the saturation throughput are depicted in Fig. \ref{fig2}, where a very good match can be observed. Furthermore, figures \ref{fig3}, \ref{fig4} and \ref{fig5} present how $\tau$, $p$ and $E[D]$ are influenced by the network size, respectively. As before, the graphs reveal an excellent match among the values produced by the proposed model and by simulation.

The higher the $UP$, the higher the transmission and collision probabilities and the higher the rate that the saturation throughput decreases as the network grows. $UP=3$, which is used to facilitate emergency traffic (e.g., emergency alerts based on vital sign monitoring), is particularly aggressive in terms of transmission probability and its performance deteriorates exponentially as the number of nodes with that type of traffic increases. The remaining $UP$'s exhibit a smoother throughput variation as the number of nodes rises and we can expect a system capacity utilization of around 40\% when the system reaches the maximum recommended number of stations (i.e., 8 nodes). 

In terms of average delay (see Fig. \ref{fig5}), traffic which belongs to $UP$s 0, 1 and 2 experiences similar performance, especially when the number of nodes remains below the recommended value of 8 nodes. On the other hand, when the network services high priority traffic, the measured average delay suffers a rapid deterioration.

$UP=0$ and $UP=1$ exhibit similar and smoother performance (Fig. \ref{fig2} and Fig. \ref{fig5}) due to their low $CP$ values (Table \ref{tab1}), which allows them to easily resolve conflict situations. $UP=2$, that has the second highest $CP_{max}$ value ($\sfrac{1}{2}$), will more frequently be involved in conflict situations, as the number of competing nodes gets larger. However, having an $r$ value of 4, will reduce its contention probability twice (see Section II and Table \ref{tab1}), in the case of consecutive collision events, before it settles to its corresponding $CP_{min}$ value. Since $CP_{min}$ of $UP=2$ is also relatively low ($\sfrac{1}{8}$), it provides it with the ability to recover from collision events. 

On the contrary, emergency $UP$ ($UP=3$) is assigned with the highest $CP$ parameters. This renders the nodes producing that kind of traffic contention-persistent. Combined with an $r$ value of 2, the transmission probability of the emergency $UP$ tends (Fig. \ref{fig3}) to rest on its assigned $CP_{min}$ value very quickly ($n>2$). The outcome, as depicted in Fig. \ref{fig4}, is a sharp increase in collision probability and, ultimately, a rapid deterioration of its throughput and delay performance. 

Based on these observations, it would not be advisable for sensor nodes, that frequently require channel access, to be configured with the highest $UP$, especially when their number becomes large. Instead, assigning to them a lower $UP$ value would be a wiser choice.
\balance

\section{Conclusions}
This paper investigated the throughput and delay performance of the slotted Aloha channel access method defined in the ETSI SmartBAN standard under saturation conditions. The analysis utilized a two-dimensional Discrete Time Markov Chain to model the performance of the protocol which effectively captures the dynamics of the slotted Aloha version of SmartBANs. The accuracy of the proposed model is validated by means of simulation and both theoretical and simulation results exhibit a very good match. The proposed model can be used to estimate the maximum performance capabilities, in terms of saturation throughput and average packet end-to-end delay, of the slotted Aloha protocol for the different user priorities specified by the standard. Future work will include the modification of the proposed model to assess the performance of slotted Aloha in heterogeneous SmartBANs (i.e., network with traffic belonging to different $UP$s) and in the presence of non-ideal channel conditions. Moreover, the capability of the protocol to efficiently support different use-cases of the SmartBAN technology will also be investigated.


\begin{thebibliography}{00}
\bibitem{b1} D. M. G. Preethichandra, L. Piyathilaka, U. Izhar, R. Samarasinghe and L. C. De Silva, \enquote{Wireless body area networks and their applications—A Review,} \emph{IEEE Access}, vol. 11, pp. 9202-9220, Jan. 2023. 
\bibitem{b2} M. Osama et al., \enquote{Internet of medical things and healthcare 4.0: Trends, requirements, challenges, and research directions,} \emph{Sensors}, vol. 23, no. 17, p. 7435, Aug. 2023.
\bibitem{b3} \emph{IEEE Standard for Local and Metropolitan Area Networks - Part 15.6: Wireless Body Area Networks}, IEEE Standard 802.15.6. Feb. 2012. [Online]. Available: \nolinkurl{https://ieeexplore.ieee.org/document/6161600}
\bibitem{b4} \emph{Smart Body Area Networks (SmartBAN); System Description}, ETSI TR 103 394 V1.1.1, Jan. 2018. [Online]. Available: \nolinkurl{https://www.etsi.org/deliver/etsi_tr/103300_103399/103394/01.01.01_60/tr_103394v010101p.pdf}
\bibitem{b5} \emph{Smart Body Area Networks (SmartBAN); Low Complexity Medium Access Control (MAC) for SmartBAN}, ETSI TR 103 325 V.1.2.1, July 2022. [Online]. Available: \nolinkurl{https://www.etsi.org/deliver/etsi_ts/103300_103399/103325/01.02.01_60/ts_103325v010201p.pdf}
\bibitem{b6} T. Paso \emph{et al.}, \enquote{An overview of ETSI TC SmartBAN MAC protocol,} \emph{in Proc. 9th Int. Symp. Med. Inf. Commun. Technol. (ISMICT)}, Mar. 2015,
pp. 10–14.
\bibitem{b7} \emph{Smart Body Area Networks (SmartBAN); Applying SmartBAN MAC (ETSI TS 103 325) for various use-cases}, ETSI TR 103 711 V.1.1.1, Oct. 2020. [Online]. Available: \nolinkurl{https://www.etsi.org/deliver/etsi_tr/103700_103799/103711/01.01.01_60/tr_103711v010101p.pdf}
\bibitem{b8} M. S. Chowdhury, K. Ashrafuzzaman and K. S. Kwak, \enquote{Saturation throughput analysis of IEEE 802.15.6 slotted ALOHA in heterogeneous conditions,} \emph{IEEE Wireless Commun. Lett.}, vol. 3, no. 3, pp. 257-260, Jun. 2014.
\bibitem{b9} T. Benmansour, T. Ahmed, S. Moussaoui and Z. Doukha, \enquote{Performance analyses of the IEEE 802.15.6 Wireless Body Area Network with heterogeneous traffic,} \emph{J. Netw. Comput. Appl.}, vol. 163, p. 102651, Aug. 2020.
\bibitem{b10} R. Khan, M. M. Alam and M. Guizani, \enquote{A flexible enhanced throughput and reduced overhead (FETRO) MAC protocol for ETSI SmartBAN,} \emph{IEEE Trans. Mobile Comput.}, vol. 21, no. 8, pp. 2671-2686, Aug. 2022.
\bibitem{b11} R. Khan and M. M. Alam, \enquote{SmartBAN performance evaluation for diverse applications,} in \emph{Proc. EAI Int. Conf. Body Area Netw.}, Florence, Italy, 2019, pp. 239-251.
\bibitem{b12} G. Bianchi, \enquote{Performance analysis of the IEEE 802.11 distributed coordination function,} \emph{IEEE J. Sel. Areas Commun.}, vol. 18, no. 3, pp. 535-547, Mar. 2000.
\bibitem{b13} B. Krishnamachari. \emph{AlohaSim}. (2021). GitHub. Accessed: Aug. 8, 2024. [Online] Available: \nolinkurl{https://github.com/ANRGUSC/AlohaSim}
\end{thebibliography}
\end{document}